\documentclass[12pt,a4paper,final]{iopart}
\usepackage{iopams}  
\usepackage[breaklinks=true,colorlinks=true,linkcolor=blue,urlcolor=blue,citecolor=blue]{hyperref}
\usepackage[dvips]{graphicx}
\usepackage{bm}

\usepackage[latin9]{inputenc}
\usepackage{textcomp}

\makeatother
\begin{document}

\title{Ginzburg-Landau theory for the magnetic and structural transitions 
in La$_{1-y}$(Ca$_{1-x}$Sr$_{x}$)$_{y}$MnO$_{3}$}

\author{L. M. Le\'on Hilario}
\address{Laboratorio de Materiales Nanoestructurados, Facultad de Ciencias, Universidad Nacional de Ingeniería, Avenida Túpac Amaru 210, Rímac 025, Lima, Perú}
\author{A.~A.~Aligia}
\address{Centro At\'{o}mico Bariloche and Instituto Balseiro, Comisi\'{o}n Nacional
de Energ\'{\i}a At\'{o}mica, CONICET, 8400 Bariloche, Argentina}

\begin{abstract}
We present a phenomenological theory for the ferromagnetic transition temperature, the magnetic susceptibility 
at high temperatures, and the structural distortion in the La$_{1-y}$(Ca$_{1-x}$Sr$_{x}$)$_{y}$MnO$_{3}$ system. 
We construct a Ginzburg-Landau free energy that describes the magnetic
and the structural transitions, and a competition between them.
The parameters of the magnetic part of the free energy are derived from a mean-field solution of
the magnetic interaction for arbitrary angular momentum. 
The theory provides a qualitative description of the observed magnetic and structural phase transitions
as functions of Sr-doping level ($x$) for $y=0.25$.

\end{abstract}

\maketitle

%\date{\today }

\section{Introduction}

\label{intro}

The compounds with colossal magnetoresistance (CMR) manganites, such as La$_{1-y}$Ca$_y$MnO$_{3}$ 
and La$_{1-y}$Sr$_y$MnO$_{3}$, have attracted a great deal interest 
in the last decades due to their rich fundamental physics and great potential to applications in 
magnetic memories and spintronics \cite{pre, sala, gold, dago, hag}. 
These compounds  are strongly correlated electron systems with strong interplays among  the  charge,  spin,  orbital,  
and lattice degrees of freedom. In particular superexchange interactions combined with Hund rules 
that favor ferromagnetic order compete with Jahn-Taller-type electron-lattice distortions,  
leading to complex electronic, magnetic, and  structural phase diagrams 
\cite{yuku,ueha,yoshi,good,mill,pra}. 

The CMR is a property of some materials, mostly manganese-based perovskite oxides, 
that enables them to dramatically change their electrical resistance in the presence of a 
magnetic field \cite{jin,mora}. 
Switching on a magnetic field helps to align neighboring spins, particularly 
near the the magnetic transition, where it is easier to line up the spins,
and in this case the resistance is lower, because hopping from Mn$^{+3}$ to neighboring Mn$^{+4}$ is 
facilitated.
The interaction responsible for the magnetic order is the double-exchange \cite{dege}. 
Specifically, in  a nearly cubic environment, the Mn$^{+4}$ ions have each $t_{2g}$ orbital occupied by one electron
forming a spin 3/2 (due to Hund rules) and the $e_g$ orbitals empty. The Mn$^{+3}$ iones have an additional electron in 
the more mobile $e_g$ orbitals forming total spin 2. This additional electron can move freely between parallel spin 3/2,
while if the $t_{2g}$ spins are disordered, the $e_g$ electrons suffer a scattering mechanism due to partial
frustration of the Hund rules. 
The double exchange also plays an 
important role in related perovskites, such as RuSr$_{2}$(Eu,Gd)Cu$_{2}$O$_{8}$ \cite{alia,gusm}.

In some CMR compounds, like (La,Ca)MnO$_3$ a coexistence of magnetic phases has been found,
which is absent in others, like (La,Sr)MnO$_3$. This is likely to be related by the fact that 
the Ca compound presents a first order magnetic phase transition, while the transition
is of second order in the Sr compound \cite{mira1,mira2}. 
In the series La$_{0.67}$(Ca$_{1-x}$Sr$_{x}$)$_{0.33}$MnO$_{3}$ a relationship between first or second order 
magnetic phase transition and the crystalline structure has been found \cite{riva}. 
The crystalline structure depends on the Sr content. At room temperature, in absence of Sr ($x=0$), 
it is orthorhombic, while in absence of 
Ca ($x=1$) is rhombohedral.  
However, the origin of the first order magnetic transition is still unclear.

In a more recent work, the phase transitions in the La$_{1-y}$(Ca$_{1-x}$Sr$_{x}$)$_{y}$MnO$_{3}$ were 
experimentally studied for 
$ 0.23 \leq y \leq 0.45$ \cite{aleja}. 
In particular a phase diagram as a function of Sr content $x$ for $y=0.25$ was constructed 
(Fig. 8 of Ref. \cite{aleja}, corresponding to Fig. 1 in this work).  The replacement 
of Ca$^{2+}$ ions with larger Sr$^{2+}$ ones reduces the critical temperature of the structural transition 
from a high temperature rhombohedral phase to a low temperature orthorhombic phase. In addition,
there is a magnetic transition from a high-temperature paramagnetic insulating phase to a low-temperature
ferromagnetic metallic phase. The structural and magnetic transitions cross as a function of Sr content,
in such a way that for low (high) $x$ the magnetic transition temperature is lower (higher) and for $x$
near 1, the structural transition disappears. In addition, for low $x$, the magnetic susceptibility
$\chi$ has a kink at the structural transition and $\chi^{-1}$ changes slope 
(Fig. 4 of Ref. \cite{aleja}, corresponding to Fig. 2 in this work). 

Motivated by these experimental results, in the present work we propose a simple Ginzburg-Landau free energy 
to describe both, the magnetic and structural transitions and also the kink in the magnetic susceptibility.
Ginzburg-Landau functionals have been useful in different contexts. For example recently one of these functionals 
were used to show induced spin-triplet pairing at the interface between an antiferromagnetic state and a singlet
superconductor \cite{alme}.
Our results provide a semiquantitative explanation of the phase diagram and the magnetic susceptibility 
in the paramagnetic phase.

The paper is organized as follows. The free-energy functional is described in section \ref{ener}. In section
\ref{compa} we discuss the parameters of the free energy by comparison with experiment. In particular 
we discuss the form of the magnetic part of the free energy for arbitrary angular momentum. Section \ref{pd} 
contains the resulting phase diagram. In section \ref{ms} we show the dependence of the magnetic susceptibility 
with temperature for some values of $x$, as in the experimental work \cite{aleja}. Section \ref{sum} contains 
a summary.

\section{Free Energy }
\label{ener}

In order to describe a magnetic and structural transitions, we propose a Ginzburg-Landau functional in terms of 
two dimensionless order parameters: $m=M/M_0$, where $M$ is the magnetization and $M_0$ the saturation magnetization
and $\delta$, proportional to the structural deformation (it is defined more precisely in section \ref{stru}).
As usual, we expand the free 
energy in terms of the two order parameters around $m=0$ and $\delta=0$ and 
consider the lowest order terms allowed by symmetry. The free energy can be written as
\begin{eqnarray}
F &=&-mh+A_{m}(T-T_{m})m^{2}+B_{m}m^{4} \nonumber
\\&&+A_{\delta}(T-T_{\delta})\delta^{2}+B_{\delta}\delta^{4}+Cm^{2}\delta^{2},   \label{f}
\end{eqnarray}
where the first three terms, with $A_{m}>0$ and $B_{m}>0$ determined below, correspond to the standard magnetic free energy 
in presence of a magnetic field $h$. The fourth and fifth term, with $A_{\delta}>0$ and $B_{\delta}>0$, 
correspond to the structural transition. Finally, the last term, 
corresponds to the coupling between magnetic and structural degrees of freedom. 
As we shall show, the observed phase diagram indicates that $C>0$. This means that the magnetization inhibits the 
structural distortion and vice versa. 
Note that $T_{m}$ ($T_{\delta}$) is the critical temperature of the magnetic (structural) transition
in absence of distortion (magnetization).

In our functional we neglect gradient terms and intergradient interactions. These are naturally important in the 
presence of inhomogeneities due to compositional changes that may be present in the system \cite{asker}. 
However, as we show, 
the main aspects of the observed phase diagram can be explained without including these terms.

\section{Determination of the coefficients of the  Ginzburg-Landau free energy}
\label{compa}

\subsection{Coefficients of the purely magnetic part}
\label{magpa}

In the mean-field approximation for a Heisenberg model with ferromagnetic nearest-neighbor interaction $I$, 
the magnetic free energy can be written as
\begin{equation}
 F_{m}=-\frac{Izm^{2}}{2(g\mu_{B})^{2}}-TS, \label{fm}
\end{equation}
where  $z$ is the number of nearest neighbors, $S$ is the entropy, and $T$ is the temperature. 
For models that include more interactions, $Iz$ should be replaced by an appropriate sum including further neighbors.
In any case, the value of this term can be obtained from the observed magnetic critical temperature and the details
of the model do not affect our treatment.

Now we discuss the form of the entropy for a magnetic system composed of ions with angular momentum $j$.
Later we discuss the specific case of our system.

To obtain the entropy as a function of $m$ for a given $j$, we proceed as follows. In mean field, 
the magnetization is given by the Brillouin function $B_j(\tilde{h}/T)$, where $\tilde{h}=g \mu_B h_{\rm eff}$ 
is proportional to  effective magnetic field $h_{\rm eff}$ acting on a particular site, 
which includes the external field $h$ 
and the effect of the ordered magnetic moments of its neighbors proportional to $m$. In addition, 
also the entropy in the mean-field approximation is given by the single-site entropy as a function
of $\tilde{h}/T$. The idea is to invert $m$ as a function of $\tilde{h}$ to obtain $S$ as a function of $m$ 
up to fourth order in $m$.

Specifically expanding the corresponding functions in a Taylor series around of $\tilde{h}=0$ one has
\begin{eqnarray}
m \approx \alpha \tilde{h}+\beta \tilde{h}^{3}  \label{mh} \\
S \approx S_0 + \gamma \tilde{h}^{2}+\zeta \tilde{h}^{4}, \label{sh}
\end{eqnarray}
where the coefficients $\alpha$, $\beta$, $\gamma$ and $\zeta$ are given below. 
Inverting Eq.  (\ref{mh}) one obtains $\tilde{h}$ as a function of the magnetization
\begin{equation}
\tilde{h}\approx \eta m+\nu m^{3}. \label{he}
\end{equation}
Replacing $m$ from Eq. (\ref{mh}) in the Eq. (\ref{he}) we obtain $\eta=1/\alpha$ and $\nu=-\beta/\alpha^{4}$. 

The desired expansion of $S$ as a function of the magnetization
\begin{equation}
S\approx S_0+\epsilon m^{2}+\xi m^{4}, \label{sm}
\end{equation}
is obtained replacing $\tilde{h}$ from Eq. (\ref{he}) in Eq. (\ref{sh}) and matching with  Eq. (\ref{sm}). 
We obtain $\epsilon=\gamma/\alpha^{2}$ and $\xi=\frac{-2\beta \gamma}{\alpha^{5}}+\frac{\zeta}{\alpha^{4}}$. Thus $S$ in terms of the magnetization becomes
\begin{equation}
S=\frac{\gamma}{\alpha^{2}} m^{2}+(\frac{-2\beta \gamma}{\alpha^{5}}+\frac{\zeta}{\alpha^{4}}) m^{4}, \label{smf}
\end{equation}

To calculate the values of the coefficients $\alpha$, $\beta$, $\gamma$ and $\zeta$, 
we use well known expressions for the statistical theory of magnetism. The magnetization 
is expressed in terms of the Brillouin function $B_{j}$
\begin{equation}
m\equiv\frac{M}{M_{0}}=B_{j}(\tilde{h}/T)=\frac{(2j+1)}{2j}\coth\left[(\frac{2j+1}{2})
\frac{\tilde{h}}{T}\right]-\frac{1}{2j}\coth\left[\frac{\tilde{h}}{2T}\right], 
\label{bri}
\end{equation}
Expanding the Brillouin function around $h=0$, and truncating it to third order, the resulting magnetization reads
\begin{equation}
m=\frac{(1+j)}{3T} \tilde{h} -\frac{(1+3j+4j^{2}+2j^{3})}{90T^{3}}\tilde{h}^{3} \;\;  \approx \alpha \tilde{h}+\beta \tilde{h}^{3}\label{briex}
\end{equation}
This result allows us to obtain the values of $\alpha$ and $\beta$:
\begin{eqnarray}
\alpha=\frac{(1+j)}{3T}   \\
\beta=-\frac{(1+3j+4j^{2}+2j^{3})}{90T^{3}}. 
\end{eqnarray}

It remains to calculate $\gamma$ and $\zeta$. 
From the definition of the Helmholtz free energy (we use units with the Boltzmann constant $k_B=1$)
\begin{equation}
F=-T\ln (Z)
\end{equation}
The entropy is given by
\begin{equation}
S=-\frac{\partial F}{\partial T}=\ln(Z) +T\frac{1}{Z}\frac{\partial Z}{\partial T}
\end{equation}
which is expressed in terms of the partition function $Z$
\begin{equation}
Z=\displaystyle \sum_{m=-j}^{j} e^{m\tilde{h}/T}.
\end{equation}
Expanding $S$ around $\tilde{h}=0$, and truncating it to fourth order, the resulting entropy can be written as
\begin{equation}
S={\rm ln}(2j+1)-\frac{j(1+j)}{6T^{2}} \tilde{h}^{2} +\frac{j(1+3j+4j^{2}+2j^{3})}{120T^{4}}\tilde{h}^{4} \;\;  \approx \gamma \tilde{h}^{2}+\zeta \tilde{h}^{4}\label{sej}
\end{equation}
Thus we obtain the values of the numbers $\gamma$ and $\zeta$
\begin{eqnarray}
\gamma=-\frac{j(1+j)}{6T^{2}}   \\
\zeta=\frac{j(1+3j+4j^{2}+2j^{3})}{120T^{4}}.
\end{eqnarray}
$S_0={\rm ln}(2j+1)$ is independent of $m$ and will be dropped in what follows.
Replacing $\alpha$, $\beta$, $\gamma$ and $\zeta$ in the Eq. (\ref{smf}) and then in the Eq. (\ref{fm}), 
we obtain finally a expression to the magnetic free energy 
\begin{equation}
F_{m}=\frac{3}{2}\frac{j}{(1+j)}(T-T_{m})m^{2}+\frac{9}{40}\frac{j(1+2j+2j^{2})}{(1+j)^{3}}Tm^{4}, \label{fmg}
\end{equation}
where $T_m=Izj(j+1)/3$, is the usual mean-field expression for the magnetic transition temperature.
This last result is a general expression for arbitrary  $j$. Note that for $j=1/2$ the result is in agreement 
with previous results \cite{bean,bust}. The linear term in $T-T_{m}$ is also present in the theory of Ref. \cite{blois}.  
From Eq. (\ref{fmg}), the parameters $A_{m}$ and $B_{m}$ of the magnetic part of the free energy Eq. (\ref{f})
are determined to be
\begin{eqnarray}
A_{m}&=& \frac{3}{2}\frac{j}{(1+j)} \label{am}\\
B_{m}(T)&=&\frac{9}{40}\frac{j(1+2j+2j^{2})}{(1+j)^{3}}T =B_{m}^{\prime}T\label{bm}
\end{eqnarray}

For our system with $y=0.25$, there is a mixture of 75\% of Mn$^{+3}$ and 25 \% of Mn$^{+4}$ with total spin 
2 and 3/2 respectively. Since the angular momentum is quenched in the system, we take 
the values of $A_m$ and $B_m$ from the corresponding average (0.75 the values for $j=2$ plus 0.25 the values 
for $j=3/2$). Therefore we obtain
 \begin{eqnarray}
A_{m}&=& 0.975 \label{amf}\\
B_{m}^{\prime}&=&0.208 \label{bmf}
\end{eqnarray}

\subsection{Transition temperatures}
\label{tt}

As discussed in the introduction, the temperatures for both transitions depend of the Sr content. 
The same happens with the ``bare'' transition temperatures $T_m$ and $T_\delta$, which correspond to a system with 
no coupling between magnetic and structural degrees of freedom ($C=0$). For small $x$ one has $T_\delta> T_m$. In this case, for $h=0$, 
using Eq. (\ref{f}), as the temperature is lowered, the first transition found is the structural one 
at the bare critical temperature $T_\delta$. Al lower temperatures, the order parameter $\delta$ of the 
structural distortion is different from zero. For intermediate temperatures 
$T_{m}^{\prime} < T < $ $T_\delta$, where $T_{m}^{\prime}<T_m$  is the real magnetic
transition temperature, the value of $\delta$ can be obtained minimizing the free energy of Eq. (\ref{f})
for $m=h=0$ with respect to $\delta^2$. This leads to 
\begin{equation}
\left .\frac{\partial F}{\partial \delta^{2}}\right|_{m=0}=A_{\delta}(T-T_{\delta})+2B_{\delta}\delta^{2}=0 \;\; 
\rightarrow \delta^{2}=\frac{A_{\delta}(T_{\delta}-T)}{2B_{\delta}}
\label{delt}
\end{equation}
For temperatures greater than or equal to the magnetic transition temperature ($T > T_{m}^{\prime}$), we can 
replace the above obtained value of $\delta$ (valid only for $m=0$) in the free energy. 
The resulting expression depends only on $m$ and $T_{m}^{\prime}$ can be obtained as usual from the
requirement that the coefficient of $m^2$ vanishes. For $m \rightarrow 0$
\begin{equation}
\frac{\partial F}{\partial m^{2}}=A_{m}(T_{m}^{\prime}-T_{m})+C\frac{A_{\delta}(T_{\delta}-T_{m}^{\prime})}{2B_{\delta}}=0 
\end{equation}
From this condition, the magnetic transition temperature is
\begin{equation}
T_{m}^{\prime}=\frac{T_{m}-\frac{A_{\delta}C}{2A_{m}B_{\delta}}T_{\delta}}{1-\frac{A_{\delta}C}{2A_{m}B_{\delta}}}. 
\label{tmp}
\end{equation}

One can proceed in a similar way for the case in which $T_\delta < T_m$. In  this case, the magnetic transition 
temperature takes place at the ``bare'' transition temperature $T_m$ and the structural transition temperature is 
$T_{\delta}^{\prime} < T_\delta $. The analysis is the same as before with corresponding parameters for structural and 
magnetic transitions interchanged.
The result is
\begin{equation}
T_{\delta}^{\prime}=\frac{T_{\delta}-\frac{A_{m}C}{2A_{\delta}B_{m}}T_{m}}{1-\frac{A_{m}C}{2A_{\delta}B_{m}}} 
\label{tdp}
\end{equation}
In this case however, since $B_{m}=B_{m}^{\prime} T_{\delta}^{\prime}$ [see Eq.  (\ref{bm})], $T_{\delta}^{\prime}$ 
enters also the second member of Eq.  (\ref{tdp}) leading to a quadratic equation in $T_{\delta}^{\prime}$.
This is discussed below.

\subsection{Structural coefficients of the free energy}
\label{stru}

Here we determine some relations between the different parameters of the free energy that come from experiment.

From the fit of the experimental data for the highest transition temperature 
as a function of the Sr concentration $x$ presented in the Ref. \cite{aleja} we obtain
\begin{eqnarray}
T_{\delta}=(712.06-668.60 x) {\rm \; K}  \label{td} \\
T_{m}=(284.15+59.86 x) {\rm \; K}        \label{tm}
\end{eqnarray}
The magnetic transition temperature given by Eq. (\ref{tmp}) can be written in the form
\begin{equation}
T_{m}^{\prime}=\frac{T_{m}-YT_{\delta}}{1-Y} \;\; \rightarrow \; Y=\frac{T_{m}^{\prime}-T_{m}}{T_{m}^{\prime}-T_{\delta}} \label{ytmp}
\end{equation}
with
\begin{equation}
Y=\frac{A_{\delta}C}{2A_{m}B_{\delta}}
\label{y}
\end{equation}
From the experimental data of the Fig. 8 in Ref. \cite{aleja}  for $x=0 \rightarrow T_{m}^{\prime} \simeq 230$ K . 
Using this value  and Eqs. (\ref{td}), (\ref{tm}) and (\ref{ytmp}) we obtain
\begin{equation}
Y=0.112 
\label{cte1}
\end{equation}
In a similar way the structural transition temperature given by Eq. (\ref{tdp}) can be written as
\begin{equation}
T_{\delta}^{\prime}=\frac{T_{\delta}-Y^{\prime}T_{m}}{1-Y^{\prime}} \;\; \rightarrow \; 
Y^{\prime}\equiv\frac{Z}{T_{\delta}^{\prime}}=\frac{T_{\delta}^{\prime}-T_{\delta}}{T_{\delta}^{\prime}-T_{m}} 
\label{yptmp}
\end{equation}
with
\begin{equation}
Z=\frac{A_{m}C}{2A_{\delta}B_{m}^{\prime}}.
\end{equation}
Solving Eq. (\ref{yptmp}), the structural transition temperature can be expressed as
\begin{equation}
T_{\delta}^{\prime}=\frac{Z+T_{\delta}}{2}+\sqrt{\left( \frac{Z+T_{\delta}}{2}\right)^{2}-ZT_{m}} \label{ytdp}
\end{equation}
Replacing the experimental values $x \simeq 0.7, T_{\delta}^{\prime} = 200$ K  of the Ref. \cite{aleja} in 
Eq. (\ref{td}) and Eq. (\ref{tm}) and then in the Eq. (\ref{yptmp}) we obtain
\begin{equation}
Z=69.9 {\rm \; K} \label{cte2}
\end{equation}

Without loss of generality, we can assume that $\delta$ is normalized to 1 for the largest
possible value at $x=0$, $T=0$ in the absence of magnetization. In this case, Eq. (\ref{delt}) leads to
\begin{equation}
1=\frac{A_{\delta}T_{\delta0}}{2B_{\delta}}, \label{cte3}
\end{equation}
where $T_{\delta0}=T_{\delta}(x=0)=712.06$ K.
Finally, from the combination of  Eqs. (\ref{amf}), (\ref{bmf}), (\ref{y}), 
(\ref{cte1}), (\ref{cte2}) and (\ref{cte3}) we obtain
\begin{eqnarray}
A_{\delta}&=& 2.563   \\
B_{\delta}&=& 912.6   {\rm \; K}   \\
C&=&  77.61 {\rm \; K}
\end{eqnarray}

\section{Phase Diagram}
\label{pd}

With the coefficients obtained above we can calculate the magnetic and structural transition temperatures
and compare them with the experimental results.

\begin{figure}[tbp]
\includegraphics[width=\columnwidth]{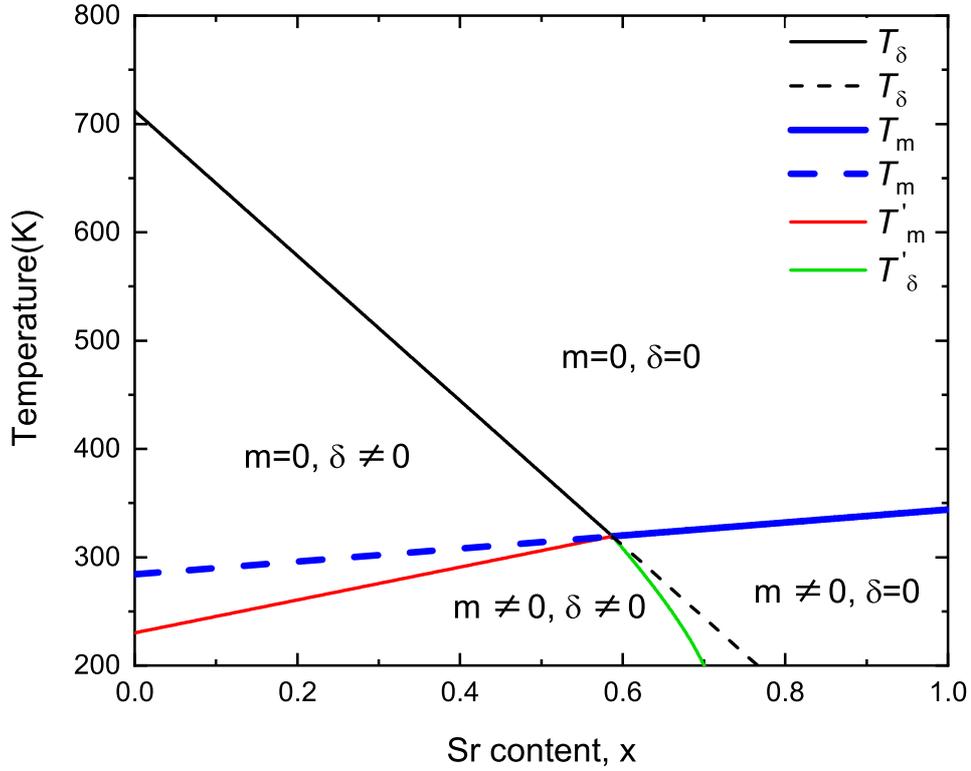}
\caption{Phase diagram Temperature vs Sr concentration $x$. Dashed lines correspond to the 
transition tempearures for $C=0$.}
\label{ctf}
\end{figure}

In  Fig. \ref{ctf} we present the  phase diagram, temperature versus Sr concentration $x$, 
for the free energy described by Eq. (\ref{f}) and the parameters obtained as described in the previous section.
The four regions of the phase diagram are characterized by the vanishing or not of the 
magnetic ($m$) and structural ($\delta$) order parameters. At high temperatures one has
$m = 0$ and $\delta =0$ corresponding to the paramagnetic rhombohedral phase. For low enough temperatures and 
low Sr content the system is in the ferromagnetic orthorhombic phase with $m \neq 0$ and $\delta\neq 0$.
These two phases meet at a tetracritical point with the paramagnetic orthorhombic phase ($m=0$ and $\delta\neq 0$)
and the ferromagnetic rhombohedral one ($m \neq 0$ and $\delta =0 $).

The phase diagram is in good agreement with the experimental one (Fig. 8 of Ref. \cite{aleja}).
Although several parameters were derived from experiment, it was not obvious that the proposed free energy functional 
describes the system.

\section{Magnetic susceptibility}
\label{ms}

In this section, we discuss the magnetic susceptibility $\partial M/ \partial h$ in the region
of low Sr content, for which $T_\delta > T_m$. It has been measured experimentally \cite{aleja}. 
One expects a kink at $T=T_{\delta}$.

For $T>T_{\delta}$  we need to consider only the magnetic free energy under a magnetic field $h$, 
the first three terms of the free energy of the system Eq.(\ref{f}) 
\begin{equation}
	F=-mh+A_{m}(T-T_{m})m^{2}+B_{m}m^{4}.
\end{equation}
When $h \rightarrow 0 $, $m$ is linear with $h$. Therefore, minimizing the free energy with respect to $m$ and 
then deriving it with respect to $h$ we obtain
\begin{equation}
	\chi_0= \frac{\partial m}{\partial h} = \frac{K}{T-T_{m}}  \;\; , \;\;\;\;\;\;\;\; K=\frac{1}{2A_{m}}.
\end{equation}

For $T^{\prime}_{m}<T<T_{\delta}$ we have to add the coupling between magnetic and structural order parameters.
Since $m \rightarrow 0$ we can use Eq. (\ref{delt}) fir the value of $\delta$ and proceed in a similar way 
as above to obtain 
\begin{equation}
	\chi_0=\frac{K^{\prime}}{T-T^{\prime}_{m}}  \;\; , \;\;\;\;\;\;\;\; K^{\prime}=\frac{1}{2A_{m}-\frac{CA_{\delta}}{B_{\delta}}}
\end{equation}

\begin{figure}[tbp]
	\includegraphics[width=\columnwidth]{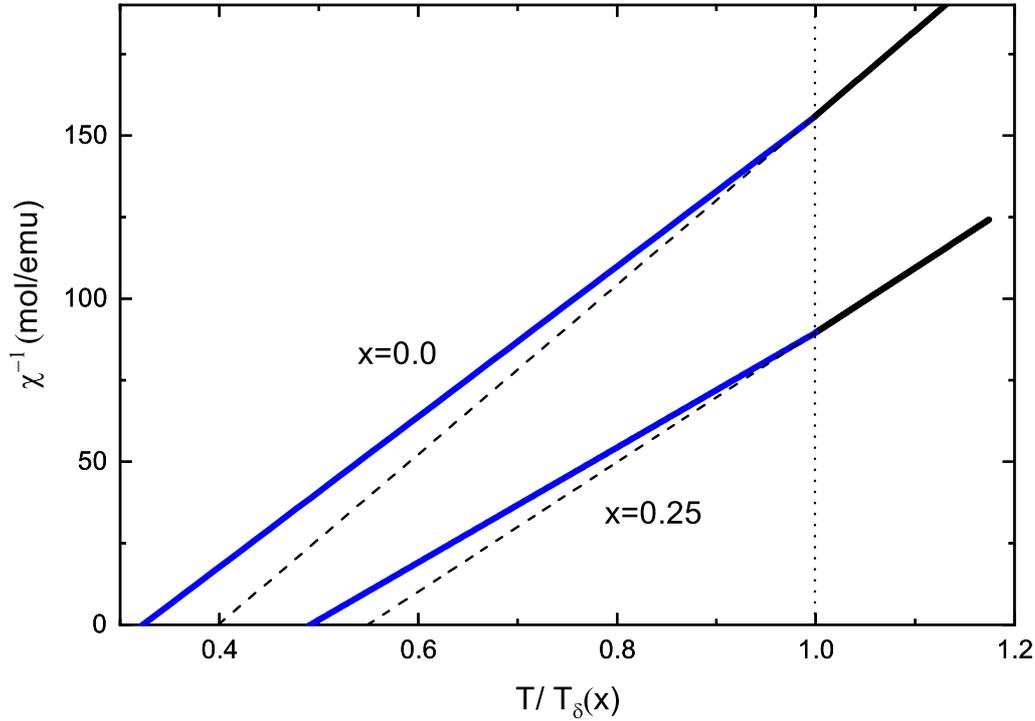}
	\caption{Inverse of the magnetic susceptibility versus $T$ for $x=0$ and $x=0.25$. 
	The dashed lines represent the continuation of the corresponding values of the high temperature 
	rhombohedral phase and the dotted line indicates the structural transition}
	\label{sus}
\end{figure}

The magnetic susceptibility is $\chi=M_0 \chi_0$. In Fig. \ref{sus} we plot $\chi^{-1}$ versus $T$ for $x=0$. 
Clearly $\chi^{-1}$ present a change of slope at the transition temperature $T_{\delta}$.
The inverse of the susceptibility for $T>T_\delta$ extrapolates to $T_m$, while for $T<T_\delta$ it 
extrapolates to $T^{\prime}_{m}$. This kink is also observed experimentally (Fig. 4 of Ref \cite{aleja}).
Experimentally a positive curvature of $\chi^{-1}$ is observed at small temperatures which is beyond the reach of our 
theory.

\section{Summary and discussion}
\label{sum}

We have presented a Ginzburg-Landau theory to describe experimentally observed magnetic and structural transitions
in La$_{1-y}$(Ca$_{1-x}$Sr$_{x}$)$_{y}$MnO$_{3}$.
For the magnetic part of the free energy, we derive the parameters from a generalization to
arbitrary angular momentum $j$ (or spin in the angular momentum is quenched) of previous results 
for $j=1/2$ \cite{bean}. This might be useful for future use of magnetic free-energy functionals.

The theory provides an explanation of the observed phase diagram and kinks in the magnetic susceptibility
vs temperature in the system for $y=0.25$ as reported by Alejandro {\it et al.} \cite{aleja}

We have considered only second-order phase transitions. For other compositions, for example $y=2/3$ \cite{riva},
first-order transitions and phase segregation was observed. Explanation of these phenomena would require
a generalization of the theory. 
First-order transitions were discussed theoretically in Ref. \cite{bust}

In any case, we believe that our results provide a qualitative understanding of the effects of the competition
between structural and magnetic order in these systems.

\section*{Acknowledgments}
We thank S. Bustingorry for useful discussions.
This work was sponsored by PIP 112-201501-00506 of CONICET and PICT
2013-1045 of the ANPCyT.  L.M.L.H is supported by the Programa Nacional
de Innovación para la Competitividad y Productividad, Innóvate Perú, research project 
INNOVATE C.389-PNICP-PIBA-2014.

\end{document}